\documentclass[aps,prb,preprint,showpacs,superscriptaddress]{revtex4}

\usepackage{indentfirst}
\usepackage{graphicx}
\usepackage{subfigure}
\usepackage{amsmath}

\begin{document}

\title{Evidence for a dynamic phase transition in [Co/Pt]$_3$ magnetic multilayers}

\author{D.T. Robb}
\email[Corresponding author: ]{drobb@berry.edu}
\affiliation{Department of Physics, Astronomy and Geology, Berry College, Mount Berry, GA 30149-5004}
\affiliation{School of Computational Science, Florida State University, Tallahassee, Florida 32306, USA}
\author{Y.H. Xu}
\altaffiliation[Permanent affiliation: ]{Department of Electrical and Computer Engineering, University of Minnesota, Minneapolis, Minnesota 55455, USA}
\affiliation{San Jose Research Center, Hitachi Global Storage Technologies, San Jose, California 95135, USA}
\author{O. Hellwig}
\affiliation{San Jose Research Center, Hitachi Global Storage Technologies, San Jose, California 95135, USA}
\author{J. McCord}
\affiliation{IFW Dresden - Institute for Metallic Materials, Postfach 270 016, 01171 Dresden, Germany}
\author{A. Berger}
% \email{Andreas.Berger@hitachigst.com}
\affiliation{San Jose Research Center, Hitachi Global Storage Technologies, San Jose, California 95135, USA}
\affiliation{CIC Nanogune Consolider, Mikeletegi Pasealekua 56, 301, E-20009 Donostia, Spain}
\author{M.A. Novotny}
% \email{novotny@erc.msstate.edu}
\affiliation{Department of Physics and Astronomy and HPC$^2$ Center for Computational Sciences, Mississippi State University, Mississippi State, Mississippi 39762, USA}
\author{P.A. Rikvold}
% \email{rikvold@csit.fsu.edu}
\affiliation{School of Computational Science, Florida State University, Tallahassee, Florida 32306, USA}
\affiliation{Center for Materials Research and Technology and Department of Physics, Florida State University, Tallahassee, Florida 32306, USA}
\affiliation{National High Magnetic Field Laboratory, Tallahassee, Florida 32310, USA}

\date{\today}

\begin{abstract}

A dynamic phase transition (DPT) with respect to the period $P$ of an applied alternating magnetic field has been observed previously in numerical simulations of magnetic systems. However, experimental evidence for this DPT has thus far been limited to
qualitative observations of hysteresis loop collapse in studies of hysteresis 
loop area scaling. Here, we present significantly stronger evidence for the
experimental observation of this DPT, in a [Co(4~\AA)/Pt(7~\AA)]$_3$-multilayer
system with strong perpendicular anisotropy. 
We applied an out-of-plane, time-varying (sawtooth) field to 
the [Co/Pt]$_3$ multilayer, in the presence of a small additional constant field, $H_b$. We then measured the resulting out-of-plane magnetization time series to produce nonequilibrium phase diagrams (NEPDs) of the
cycle-averaged magnetization, $Q$, and its variance, $\sigma ^2 (Q)$, as
functions of $P$ and $H_b$. The experimental NEPDs are found 
to strongly resemble those calculated from simulations of a kinetic Ising 
model under analagous conditions. The similarity of the experimental and 
simulated NEPDs, in particular the presence of a localized peak in the 
variance $\sigma ^2 (Q)$ in the experimental results, constitutes strong 
evidence for the presence of this DPT in our magnetic multilayer samples. 
Technical challenges related to the hysteretic nature and response time of the 
electromagnet used to generate the time-varying applied field precluded us 
from extracting meaningful critical scaling exponents from the current data. 
However, based on our results, we propose refinements to the experimental 
procedure which could potentially enable the determination of critical 
exponents in the future.

\end{abstract}


\maketitle

\section{Introduction \label{s:intro}}

A dynamic phase transition (DPT), in which
the dynamical behavior of a nonequilibrium system changes in a singular way 
at a critical value of a system parameter, can provide insight into 
the often complex behavior of such systems. DPTs of various kinds
have been identified and studied in chemical, \cite{kn:woodward02}
charge-density wave, \cite{kn:ogawa02} and superconducting systems. \cite{kn:kes04}
The DPT of interest here was first identified in computer simulations of 
magnetic systems. \cite{kn:tome90} When a bistable magnetic system in its 
ferromagnetic phase is exposed to an alternating magnetic field, the response 
depends strongly on the period of the applied field. At low values of the 
field period, the system effectively cannot respond to the rapidly changing 
field, and its magnetization oscillates in a restricted range around one of 
its two nonzero (magnetized) values. At high values of the period, the 
magnetization can follow the field, resulting in a square, symmetric hysteresis
loop. This behavior suggests the cycle-averaged magnetization Q 
(with a value near $\pm 1$ at low field period, and near 0 at high 
field period) as a candidate for a `dynamic' order parameter. It has indeed 
been shown computationally \cite{kn:sides99} that at a critical period $P_c$, 
there exists a second-order phase transition with respect to $Q$, with 
critical exponents consistent with those of the equilibrium Ising transition. 
For further description of this DPT and the numerical evidence for the 
second-order phase transition, see 
Refs.~\onlinecite{kn:sides99,kn:korniss00,kn:chakrabarti99}. 

Also associated with the appearance of a non-zero value of $Q$ below the critical period $P_c$ is a significant decrease in the hystersis-loop area. Since first being identified, this DPT has received 
a great deal of attention in numerical simulations \cite{kn:fujiwara04, kn:acharyya04, kn:jang03b} 
and analytical work. \cite{kn:meilikhov04, kn:dutta04, kn:fujisaka01}  
However, to date there have been only tentative experimental 
indications of its presence, principally the collapse of the hystersis loop area with decreasing field period, in several studies of hysteresis scaling properties in magnetic thin films.\cite{kn:jiang95, kn:suen97}  

It is important to understand the extent to which this DPT, which has been well studied computationally, can
be realized in an experimental system. To achieve this, one must investigate the quantitative
behavior of the dynamic order parameter itself. 
To this end, we identified an experimental system, an ultra-thin [Co(4~\AA )/Pt(7~\AA )]$_3$-multilayer, which possesses
strong perpedicular anisotropy and which exhibits substantial similarities \cite{Ising.simplification.endnote} to a two-dimensional kinetic
Ising model, \cite{kn:hellwig03,kn:carcia85,kn:allenspach94,kn:berger95,kn:berger97,kn:yafet88,kn:mills88,kn:back95}
for which the DPT is well-documented. \cite{kn:korniss00, kn:sides99}
We present here the first quantititative data on the behavior of the dynamic order
parameter in this Ising-like experimental system, 
and we compare it to the behavior observed 
in kinetic Monte Carlo (MC) simulations. 
Recent computational work \cite{kn:robb07} by our group on the two-dimensional kinetic Ising model identified the cycle-averaged magnetic field, $H_b$, as a significant component of the field conjugate to $Q$, and established the existence of a fluctuation-dissipation relation (FDR) between the 
susceptibility $\partial Q / \partial H_b$ and the variance $\sigma^2(Q)$ in the vicinity of the critical period. Applying those results here, we
use the variance $\sigma ^2(Q)$ as a proxy for
the susceptibility $\partial Q / \partial H_b$ in our comparisons of experimental and computational results.

In order to establish the presence of a phase transition unequivocally, it is necessary to demonstrate power-law scaling with well-defined critical exponents.
While this often can be done straightforwardly enough in computational work, 
experimentally it requires precise
measurements at carefully controlled values of the relevant thermodynamic variables. This
was achieved for the equilibrium Curie transition in an experimental thin film only fairly
recently.\cite{kn:back95} In the present case of a [Co/Pt]$_3$ multilayer driven by an alternating magnetic field, the nonlinear, hysteretic nature of the electromagnet used to generate the alternating
field resulted in small fluctuations of the cycle-averaged field, $H_b$, rendering it impossible to
study the experimental system at $H_b = 0$, as done in all but the most recent 
\cite{kn:robb07} computational studies. Moreover, it was not 
possible to establish scaling relations with statistical significance from our 
data, despite achieving an applied field and bias field amplitude precision 
of 0.5\% and 0.1\%, respectively, in our measurements. Nonetheless, our data 
for $Q$ and $\sigma ^2 (Q)$ are consistent with power-law scaling near the 
critical point.

In spite of this, we argue that the similarity of the behavior of the cycle-averaged magnetization $Q$ in the multilayer to its behavior in the kinetic Ising model, coupled with an explanation of the differences which do exist in terms of known physical properties of the multilayer, provides
strong evidence for the presence of this DPT in the multilayer. In addition, we propose refinements in the experimental technique which may produce in the future the precision required to investigate power-law scaling. However, peaks in response functions also consitute evidence
for phase transitions, and our observation of a peak in the variance 
$\sigma ^2 (Q)$ represents significantly stronger evidence for this DPT in an experimental system than that previously published.\cite{kn:jiang95, kn:suen97}

The importance of the present study is twofold. First, it provides new 
insight into the dynamic behavior of an ultrathin magnetic film system
whose properties are important to the development of future generations
of ultrahigh density information-recording technologies. Second, it provides 
valuable experimental input to theorists' efforts to develop a comprehensive 
understanding of phase transitions in far-from-equililibrium systems.
 
This paper is organized as follows. In Section \ref{s:expt}, we describe
the experimental setup and procedure. In Section \ref{s:results}, we present
our experimental results, consisting of directly measured hysteresis loops and 
magnetization time series, as well as 
nonequilibrium phase diagrams (NEPDs) used to characterize the behavior of the
dynamic order parameter and its fluctuations. 
In Section \ref{s:comparison.discussion}, we compare the experimental results
to those of computer simulations of a kinetic Ising model and 
argue that the similarities provide strong evidence for the
presence of the DPT in the multilayer system. In Section \ref{s:conclusion} we present our conclusions, as well as suggestions for
further experimental work.

\section{Experimental setup \label{s:expt}}

The ultra-thin [Co(4~\AA )/Pt(7~\AA )]$_3$-multilayer samples were prepared by 
low-pressure (3~mtorr Ar) magnetron sputtering onto ambient temperature 
silicon-nitride coated Si substrates. 
We deposited 200~\AA ~ Pt as a seed layer, and the samples were 
coated with 20~\AA ~Pt to avoid contamination. These 
ferromagnetic multilayers have an easy axis along the surface 
normal and strong uniaxial anisotropy, resulting in high 
remanent magnetization and square out-of-plane hysteresis loops. 
\cite{kn:hellwig03} X-ray diffraction confirmed a 
(111) crystalline texture, with a lateral crystallographic coherence 
length from several tens to several hundreds of nm.

Due to the strong ferromagnetic interlayer coupling, mediated by the Pt between 
adjacent Co layers, \cite{kn:carcia85} the entire multilayer acts as a single magnetic film with strong uniaxial anisotropy.  While out-of-plane magnetized 
ferromagnetic films tend to form equilibrium domain structures to reduce 
magnetostatic interactions, in ultra-thin films the energy reduction resulting
from domain formation is extremely small, so that this effect is 
strongly suppressed. \cite{kn:yafet88, kn:allenspach94, kn:berger95, kn:berger97} We 
measured the effective anisotropy field of the multilayer to be 
$H_k \approx 6$~kOe, more than an order of magnitude larger than the fields used
in our experiment. The spins should therefore remain
strongly collinear perpendicular to the film.  These experimental facts,
combined with theoretical \cite{kn:mills88} and experimental
\cite{kn:back95} evidence that ultrathin films with uniaxial anisotropy are
in the same universality class as the equilibrium Ising model, suggest
that the two-dimensional kinetic Ising model can serve as a useful model for 
the {\em nonequilibrium} behavior of our multilayer. 

To study magnetization reversal in the multilayer, we measured the polar 
magneto-optical Kerr effect (MOKE), which is proportional to the 
magnetization along the surface normal. 
A time-dependent magnetic field along the surface normal
was provided by an electromagnet. In addition, for reasons discussed in Sec.~\ref{s:results},
a small constant additional magnetic field - the `bias field', $H_b$ - 
was applied by separate means. The 
total (time-varying) field actually experienced by the multilayer 
was monitored by a Hall probe. 
Data were recorded for two multilayer samples, A and B.
The electromagnet was driven by a computer-controlled bi-polar power supply to 
provide a saw-tooth field sequence for several  
values of the period between $P = 8.7$~s and $62.3$~s (sample B: between
$P=7.6$~s and $39.6$ s). For each sequence, 
we first measured two complete, saturated hysteresis loops using a 
large field amplitude $H_s = 740\pm 5 $~Oe (for both samples A and B). 
Next, in the actual
measurement sequence, the amplitude was lowered to $H_0 = 366$~Oe 
(sample B: $H_0 = 344$~Oe), and 49 identical cycles were run. By monitoring the field
sequence, we determined that the variation of $H_0$ during the 49 cycles was
less than 1\%.  Finally, the amplitude was increased back to $H_s$ to trace 
out another complete hysteresis loop. 

The two previous experiments\cite{kn:jiang95, kn:suen97} which observed
hysteresis loop collapse with decreasing field period were conducted on
3 monolayer (3 ML) Co/Cu(001) ultrathin films with in-plane uniaxial anisotropy
\cite{kn:jiang95} and on 2-3 ML Fe/W(110) ultrathin films with 
perpendicular uniaxial anisotropy.\cite{kn:suen97} These experiments also
employed an electromagnet to generate the time-varying magnetic field,
a Hall probe to record the magnetic field at the location of the sample, and
the MOKE effect to measure the sample magnetization. In our experiment, however,
two additional steps were taken to enable a more thorough investigation 
of the experimental behavior of the cycle-averaged magnetization, 
\begin{equation}
Q_i = \frac{1}{P}\int _{(i-1)P}^{iP} {m(t) dt},
\label{dpt.def}
\end{equation}
previously identified as the order parameter for the DPT in the kinetic 
Ising model.\cite{kn:sides99} 
(The index $i$ in Eq.~(\ref{dpt.def}) denotes the number of the field cycle.) 
	
First, as mentioned above, a significant component
of the field conjugate to the dynamic order parameter was identified 
in a recent computational study \cite{kn:robb07} as the cycle-averaged
value of the magnetic field, or `bias field', 
\begin{equation}
H_b = \langle H(t) \rangle = \frac{1}{P}\int _{0}^{P} {H(t) dt}.
\label{bias.field.definition}
\end{equation}
In Ref.~\onlinecite{kn:robb07}, with $H_b$ taken as a small constant field 
superposed on a time-varying square-wave 
field, the conjugate field scaling exponent $\delta_{\rm d}$ in the DPT was 
shown to agree with the equilibrium Ising field scaling exponent 
$\delta_{\rm e} = 15$ to within computational error. 
The bias field was found to have a signficant effect on 
the dynamic order parameter $Q$, especially
near the critical period where the susceptibility $\partial Q / \partial H_b$
has its maximum. 
In a computer simulation, one can study the DPT in precisely zero bias field without
difficulty. In experiment, however, due to nonlinearities in the electromagnet used to 
generate the time-varying field, the actual bias field $H_b$ experienced by the sample
will inevitably exhibit finite values, which may even fluctuate slightly during the
measurement sequence.  We therefore chose to run 
experiments at a series of different 
values of the bias field, which were measured and controlled carefully 
using the Hall probe,\cite{zero.hall.endnote} at each value of the field period
studied.

Second, a recent study of thicker [Co/Pt]$_{50}$ multilayers \cite{kn:davies04}
found that hard-to-reverse, residual bubble domains could persist beyond
an apparent saturation field, and could serve as nucleation sites during 
magnetization reversal when the field was subsequentely reversed. 
In the experiments reported here, the two complete loops
at high saturating field before the measurement sequence 
ensured that the samples began each run in a well defined magnetic state,
and that the effects of residual bubble domains at the lower field magnitude
of the measurement cycle were consistent across different data runs.
In addition, comparison of the complete loops before
and after the measurement sequence served to verify the stability of 
our experimental setup during each run.

\section{Experimental results \label{s:results}}

\begin{figure}[ht]
\begin{center}
\includegraphics[width=3.0in,angle=0]{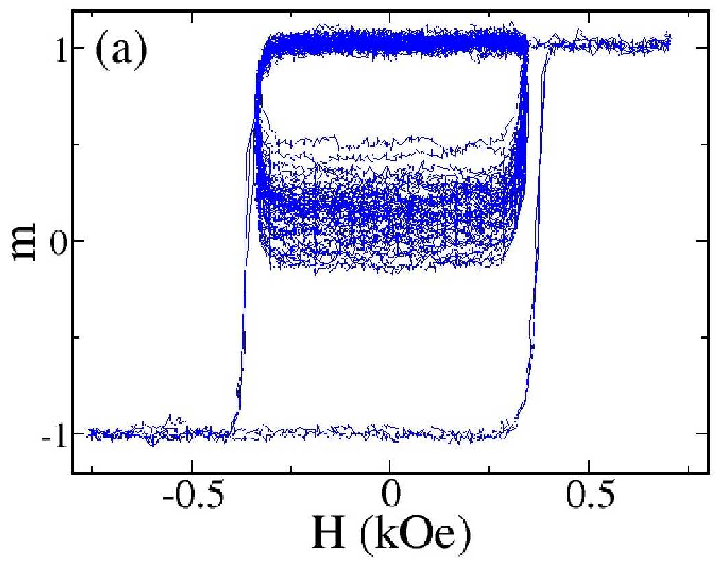} 
\includegraphics[width=3.0in,angle=0]{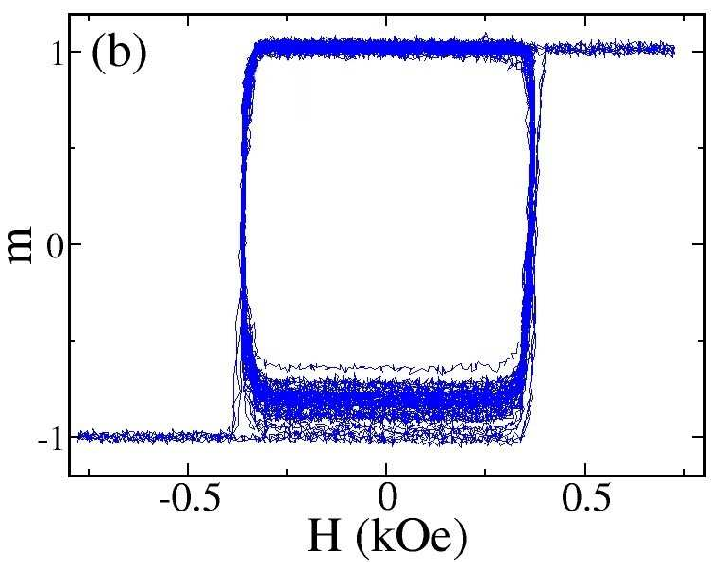} 
\end{center}
\caption
{\label{mh.data} 
Experimental data of the normalized magnetization, $m$, vs field $H$ for multilayer
sample A. Data were taken for two initial saturated loops, 49 loops at amplitude 
$H_0 = 0.366 \pm 0.003$~kOe, and one final saturated loop.  (a) Period $P = 16.2$~s. 
(b) $P = 38.1$~s. In both cases, the bias field, which is defined and discussed 
in Section~\ref{s:results}, was $H_b = +1.1 \pm 0.5$ Oe. }
\end{figure}

The hysteresis loops and magnetization time series in 
samples A and B had very similar characteristics, so we show only those 
of sample A (the NEPDs for both samples A and B will be presented, however).
Figure \ref{mh.data} shows the normalized
magnetization, $m$, vs $H$ for periods $P = 16.2$~s and $P = 38.1$~s  
in sample A.
The initial and final complete loops 
exhibit full remanent magnetization {\em in a single-domain state}, and a 
sharp reversal region. The loops in the measurement sequence, at field magnitude
$H_0$, are also square with a sharp reversal region. However, a sharp 
suppression of the magnetization reversal process is visible
below a pinning field $H_p = 288 \pm 2$~Oe (sample B: $H_p = 260 \pm
6$~Oe).\cite{pinning.method.endnote} 
The pinning phenomenon has been studied
quantitatively in multilayer films, and it is understood
to result from a distribution of energy barriers to domain wall motion
resulting from lattice defects and variations in the film thickness. 
\cite{kn:ferre04,kn:lemerle98,kn:dellatorre02,kn:chen02,kn:dellatorre98}

\begin{figure}[ht]
\includegraphics[width=4.0in,angle=0]{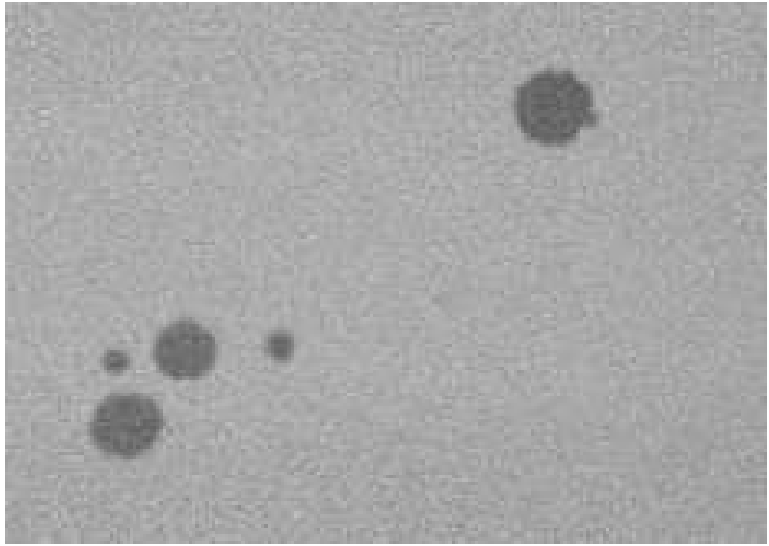} 
\caption
{\label{kerr.image} 
Kerr microscope image of magnetization in a sister [Co/Pt]$_3$ multilayer sample, taken shortly
after the start of magnetization reversal (at $m \approx 0.93$). The area represented in the image
had physical dimensions 0.88 $\times$~ 0.62 mm$^2$.}
\end{figure}

The Kerr microscope image in Fig.~\ref{kerr.image} illustrates the domain reversal 
pattern in a sister sample, at magnetization $m \approx 0.9$ after nucleation from 
a positive saturated state. It shows clearly that, above the pinning field
$H_p$, the reversal within our 1.0 mm$^2$
MOKE laser spot occurs by a multidroplet (MD) process \cite{kn:rikvold94} 
of nucleation and growth of approximately circular domains. 
This reversal pattern is similar to that previously observed in ultrathin, 
single-layer CoPt films,\cite{kn:ferre99} and can be contrasted with the
reversal pattern in thicker [Co/Pt]$_{50}$ multilayers,\cite{kn:davies04} 
in which larger magnetostatic interactions
lead to the formation of stripe domains. The MD process is quite similar to
that occurring near the DPT in the kinetic Ising model, 
\cite{kn:korniss00} indicating that, if the effect of the pinning field
is properly accounted for in the analysis, the kinetic Ising model can serve 
as a reasonable model of the multilayer. 

As described in Section \ref{s:expt}, given the important role of the bias
field $H_b$ as shown in Ref.~\onlinecite{kn:robb07},
we have studied the behavior (the average value and the variance) 
of $Q$ in the multilayer with respect to
both $P$ and $H_b$, and compared this to the corresponding behavior of 
$Q$ in the 
two-dimensional kinetic Ising model. The two previous experimental reports
of the collapse of the hysteresis loop with decreasing field period concentrated
principally on the decrease of the hysteresis loop area (cf. Fig.~1 
of Ref.~\onlinecite{kn:jiang95}, and Figs.~1 and 2 of 
Ref.~\onlinecite{kn:suen97}). 
The cycle-averaged magnetization was also plotted, in Fig.~3 of 
Ref.~\onlinecite{kn:jiang95}, 
where its magnitude was shown to become nonzero and then to increase as the
applied field amplitude is decreased at a constant value of the field period. 
However, the bias field, which has a 
strong effect on the cycle-averaged magnetization near the
dynamic phase transition in computational studies \cite{kn:robb07} 
was not carefully controlled, 
and the variance in the cycle-averaged magnetization was not recorded.

From an initial value of $H_b = -3.9$~Oe 
(sample B: $-7.4$~Oe), the bias field, as generated separately
from the time-varying field of the electromagnet, was increased in steps of
0.5 Oe (sample B: steps of $1.1$~Oe). 
The actual value of the bias field during the 49 measurement cyles was determined, using data
from the Hall probe, to fluctuate at the 0.1\% level of the applied field amplitude, $H_0$.
Specifically, the actual bias field could be characterized by a confidence interval 
$(H_b -\Delta H_b / 2, H_b + \Delta H_b / 2)$, with $\Delta H_b = 1.0$~Oe.

In Fig.~\ref{mt.data.experiment}(a), 
magnetization time series for sample A at $P = 16.2$~s 
are shown in a strong negative, weak negative,
and weak positive bias field. The corresponding time series, $Q_i$, of the 
cycle-averaged magnetization are plotted in Fig.~\ref{mt.data.experiment}(b). 
For both negative $H_b$ values, the cycle-averaged value of the magnetization 
(and in fact of any quantity) settles eventually into a non-equilibrium steady state (NESS), 
so that $Q_i$ is observed in Fig.~\ref{mt.data.experiment}(b) to fluctuate
around an average value $\langle Q \rangle < 0$. 
(The fluctuations of $Q_i$ in
the NESS arise from variations in the minor loop amplitude, as seen in
Fig.\ref{mh.data}(a), which are caused principally by 
thermally-induced
fluctuations in the extent and timing of nucleation events.) 
The transition to the NESS with $\langle Q \rangle < 0$ takes longer in 
the weak negative $H_b$ than in the strong
negative $H_b$, and it does not occur at all in the weak positive $H_b$.
At the longer period $P = 38.1$~s, in contrast, the time series 
adjust quickly to a NESS with a small $\langle Q \rangle$ value of the 
same sign as $H_b$, as shown in Figs.~\ref{mt.data.experiment}(c)
and \ref{mt.data.experiment}(d). 
This behavior will be compared to that in the kinetic Ising model in the next section. 

To characterize the behavior of the cycle-averaged magnetization in the 
multilayer more fully, we measured its average
$\langle Q \rangle$ over the field cycles for which the system was in a NESS, for 
a range of values of $H_b$ at each period $P$. We will refer to the resulting contour plot 
of $\langle Q \rangle$ vs $P$ and $H_b$, shown in Fig.~\ref{nepd.sampleA}(a), as
a non-equilibrium phase diagram (NEPD), so named in analogy to plots of 
thermodynamic relationships in equilibrium phase diagrams.  
As described in Section
\ref{s:intro}, the recent confirmation \cite{kn:robb07} of a FDR
near the critical point of the DPT in the two-dimensional kinetic Ising model 
justifies the use of the variance as a proxy for the susceptibility in
evaluating evidence for the DPT.
The NEPD of the variance $\sigma ^2(Q)$ within the NESS is
shown in Fig.~\ref{nepd.sampleA}(b). The corresponding
NEPDs for sample B are presented in Fig.~\ref{nepd.sampleB}. 
To determine in which field cycle the system should be 
considered to have entered a NESS, we used the following procedure. The first 10 field
cycles were discarded to minimize initial transient effects. We then identified
the first subsequent field cycle, $n$, for which the slope of the 
remaining values $\{ Q_i, i \ge n \}$
could not be statistically distinguished from zero. 
If more than 15 values of $Q_i$ remained, the mean
$\langle Q \rangle$ and variance $\sigma^2(Q)$ of these values were calculated; otherwise,
no data were recorded on the NEPDs for that pair of values $(P,H_b)$.
This procedure ensured that, for example for the measurement sequences
in strong negative and weak negative bias fields in 
Fig. \ref{mt.data.experiment}(a), 
the field cycles making up a transition to the NESS with
$\langle Q \rangle > 0$ were not included as part of the respective NESS's.

In Fig. \ref{nepd.sampleA}(a), the boundary between regions with negative 
$\langle Q \rangle$ and those with positive $\langle Q \rangle$ occurs, for the lower
periods, at a slightly negative bias field, $H_b \approx -1$~Oe. Similarly, the 
maximum of  $\sigma ^2(Q)$ occurs, for the lower periods, at a slightly 
negative bias field, $H_b \approx -1.5$~Oe.
This appears to contradict the expected $M-H$ inversion
symmetry of a ferromagnetic system, which predicts that the NEPDs should
be symmetric with respect to the bias field $H_b$. However, it can 
be explained by the presence of residual, hard-to-reverse bubble domains, 
with coercivities $H_c$ in the range $H_0 < |H_c| < H_s$,
as observed recently in thicker [Co/Pt]$_{50}$ multilayers.\cite{kn:davies04}
Since the final saturating field before our measurement sequence begins is
positive, such residual domains (if present) would remain positively magnetized during the
measurement sequence, in which the field magnitudes $|H(t)| < H_0$  
would be too weak to reverse them.
The bubble domains would then serve as nucleation centers for reversal on the 
increasing branch of the hysteresis loops during the measurement sequence, lowering
the nucleation field on this branch. 
This effect can indeed be observed in
Fig.~\ref{illustrate.persistent}(a), in which the nucleation field $H_{n,+}$ on 
the increasing branch (from the negatively magnetized state) in the final 
five measurement cycles is seen to be smaller than the nucleation field in 
the increasing branch of the cycles at saturation field $H_s$. In contrast,
the nucleation field $H_{n,-}$ on the decreasing branch 
(from the positive magnetized state) in the first five
measurement cycles is approximately equal to the nucleation field in the
decreasing branch of the cycles at saturated field $H_s$.
 
This asymmtery in the nucleation fields has an effect on the value of $Q$ 
in equal and opposite bias fields, $H_b$ and $-H_b$, at
a given field period in sample A. 
In the positive bias field $H_b$, the reversal window beginning from
positive saturation consists of the field intervals $(H_{n,-},-H_0+H_b)$ and
$(-H_0+H_b,-H_p)$, whereas in the negative bias field $-H_b$, the reversal
window beginning from negative saturation 
consists of the field intervals $(H_{n,+},H_0-H_b)$
and $(H_0-H_b,H_p)$. Since $H_{n,+} < -H_{n,-}$, the magnetization reversal
proceeds farther in the bias field $-H_b$ than in the bias field $+H_b$, leading
to an asymmetry in the values of the cycle-average magnetization $Q$, with
$Q|_{H_b} > -Q|_{-H_b}$. Thus, the boundary between the regions with negative
$\langle Q \rangle$ and positive $\langle Q \rangle$ is shifted from 
$H_b = 0$~Oe to the slightly negative bias field $H_b \approx -1.0$~Oe.
 
\section{Comparison to kinetic Ising model simulations \label{s:comparison.discussion}}

We now compare the experimental results to the behavior of the 
two-dimensional kinetic Ising model with nearest-neighbor
ferromagnetic interactions, for which the DPT has been conclusively established,
\cite{kn:korniss00, kn:sides99} under conditions similar
to those of the experiments. The energy of the Ising system is given by
\begin{equation}
E = - J \sum_{\langle i, j \rangle} S_i S_j -  H(t) \sum_i S_i, 
\label{Ising.energy}
\end{equation}
with $J>0$ the exchange constant, $H(t)$ the time-dependent magnetic field, and $S_i  = \pm 1$ the spin magnetic
moment at site $i$.  We use the Glauber transition rule, $P(S_i \rightarrow -S_i) = 1/(1+e^{\beta \Delta E})$, with the next spin to be updated chosen at random.  Having verified that larger 
lattice sizes do not change the results appreciably, we use
a $180 \times 180$ square lattice with periodic boundary conditions.
The Curie temperature of the multilayer was estimated from Ref. \onlinecite{kn:zeper91} 
as $T_c \approx 600$~K.  We therefore choose the simulation temperature $T = 0.5~T_c$, 
to correspond to the temperature $T \approx 300~K$ of the experiments. 
As in Ref. \onlinecite{kn:sides99}, the reversal dynamics
are found to be MD for $H_0 = 0.8J$ at this temperature, with mean reversal time 
$ \langle \tau \rangle = 59.23 \pm 0.06$ Monte Carlo steps per spin (MCSS). For a 
sawtooth waveform with amplitude $H_0 = 0.8J$, we determine the DPT by finite-size 
scaling analysis \cite{kn:korniss00, kn:sides99} to occur at $P_c = 493 \pm 2$ MCSS.  

\begin{figure}[ht]
\begin{center}
\includegraphics[width=3.0in,angle=0]{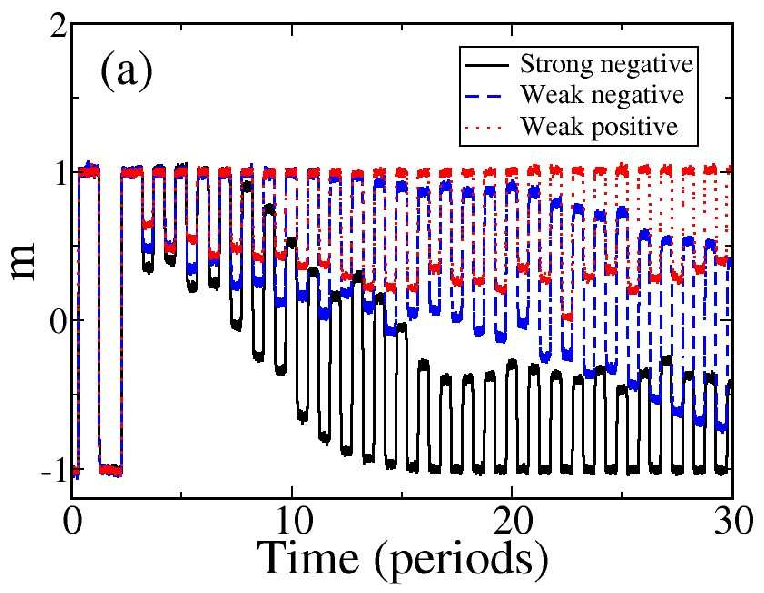} 
\includegraphics[width=3.0in,angle=0]{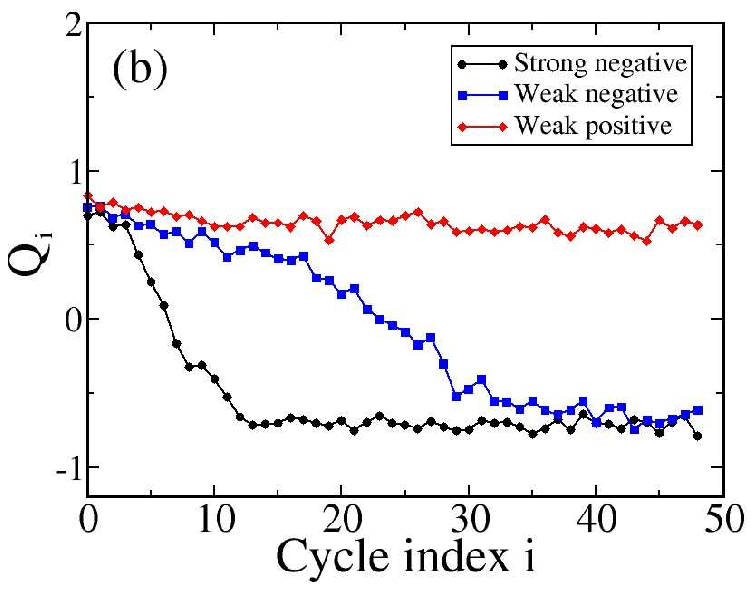} 
\end{center}
\begin{center}
\includegraphics[width=3.0in,angle=0]{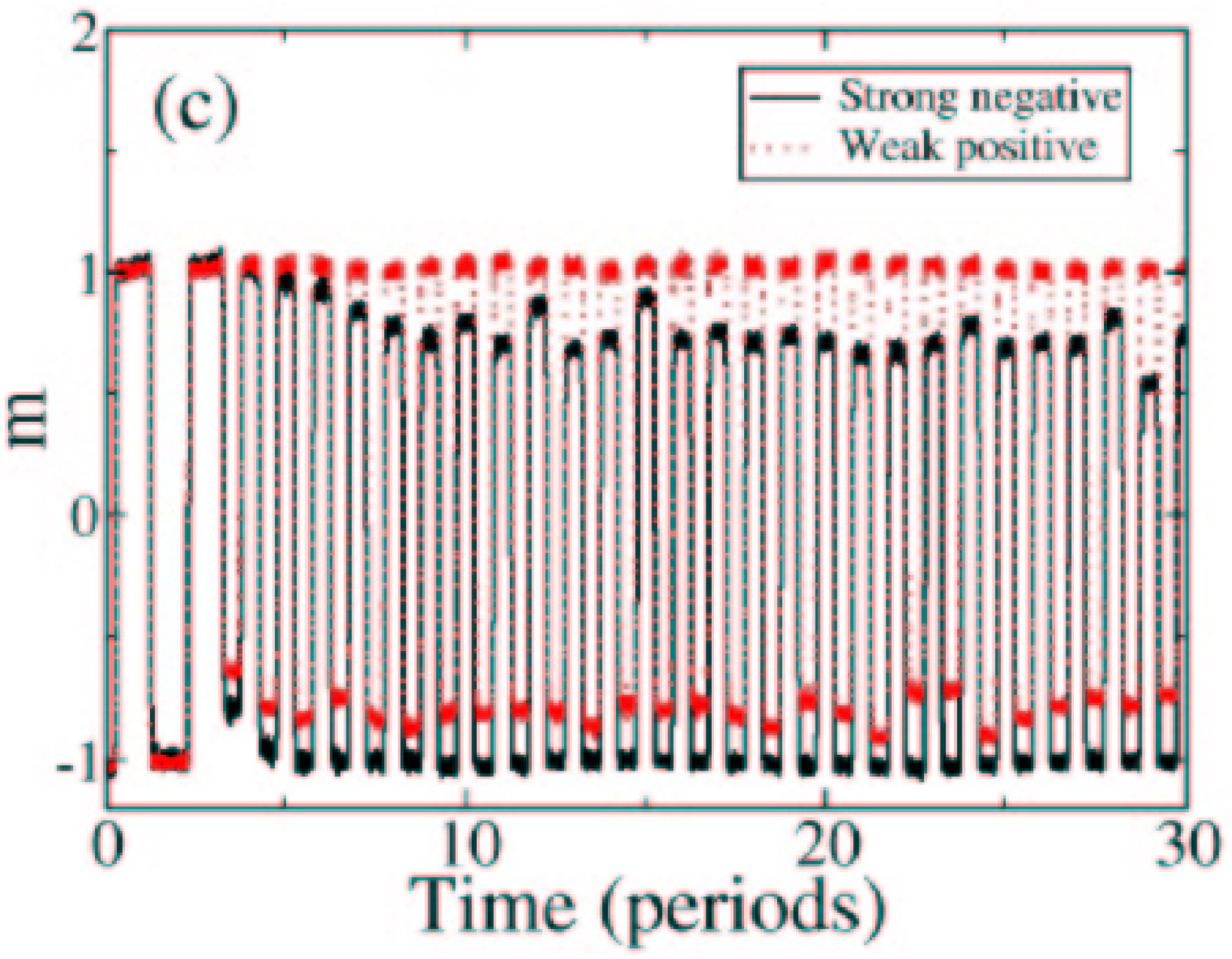} 
\includegraphics[width=3.0in,angle=0]{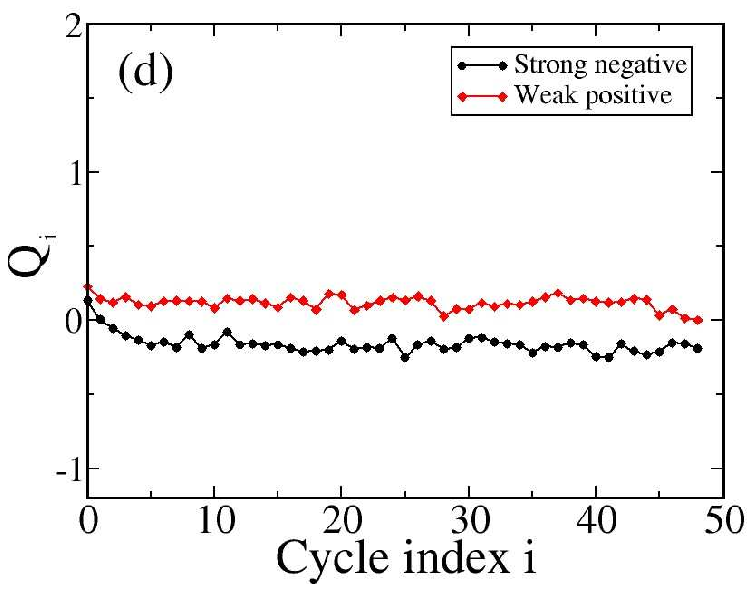} 
\end{center}
\caption
{\label{mt.data.experiment} (color online). 
Time series of the normalized magnetization, $m$, and of the cycle-averaged magnetization, $Q_i$, in multilayer sample A, at two different values of the period. 
(a) Period $P = 16.2$~s, magnetization time series. (b) $P = 16.2$~s, $Q_i$ vs 
measurement cycle index $i$. (c) $P = 38.1$~s, magnetization time series. 
(d) $P = 38.1$~s, $Q_i$ vs measurement cycle index $i$. 
Note that for clarity, only the two initial saturated cycles and the first 26 measurement cycles 
of the magnetization time series are plotted; however, the full time series of $Q_i$
(calculated from the 49 measurement cycles only) are shown.  The bias field values 
corresponding to strong negative, weak negative, and weak positive are 
$H_b = -3.3 \pm 0.5, -0.9 \pm 0.5,$ and $+1.1 \pm 0.5$~Oe, respectively.}
\end{figure}

To recreate the experimental conditions as closely as possible, we simulated
magnetization time series consisting of two saturated loops at 
$H_s = 1.6J$, followed by 50 field cycles at $H_0 = 0.8J$. The period values
were selected from both below and above $P_c$, from $P = 368$ to
$1575$~MCSS. At each value of $P$, we collected magnetization time series 
for a range of bias fields from $H_b = -0.02 H_0$ to $0.01H_0$.  
Figures~\ref{mt.data.simulation}(a) and \ref{mt.data.simulation}(b) show 
simulated magnetization 
time series, and the corresponding time series of the 
cycle-averaged magnetization, $Q_i$,
for a strong negative,  weak negative, and  strong positive 
bias field at period $P=473$~MCSS, just below $P_c=493$~MCSS. 
As in the experimental time series in Figs.~\ref{mt.data.experiment}(a) and
\ref{mt.data.experiment}(b), the transition to the NESS with $\langle Q \rangle < 0$
in the simulated time series takes longer in the weak negative $H_b$ than in the strong negative $H_b$
, and it does not occur at all in the weak positive $H_b$. In addition, the 
behavior of the simulation in strong negative and weak positive $H_b$ at the 
higher period $P=1500$~MCSS, shown in Figs.~\ref{mt.data.simulation}(c) and
\ref{mt.data.simulation}(d), 
resembles the behavior of the experimental system at higher period shown in 
Figs.~\ref{mt.data.experiment}(c) and \ref{mt.data.experiment}(d). The larger splitting
between the average values of $\langle Q \rangle$ in Fig.~\ref{mt.data.experiment}(d),
relative to Fig.~\ref{mt.data.simulation}(d), is due mainly to the effect of the
pinning field in the experimental system, which results in an increased sensitivity to
the bias field, as discussed later in this section. 

The similarity of the simulation data in Fig.~\ref{mt.data.simulation} to the 
experimental data in Fig.~\ref{mt.data.experiment}
suggests that the experimental system may be near criticality just above
$P=16.2$~s, but this similarity does not by itself provide conclusive 
evidence of a DPT.
To evaluate the question more thoroughly,  we constructed 
NEPDs of $\langle Q \rangle $ and $\sigma^2(Q)$, measured in the NESS
in simulation, with respect to $P$ and $H_b$.
Two procedures were used to generate simulation data in a NESS. 
In the first procedure, we mimicked the experimental data analysis precisely,
by identifying the part of the simulated time series 
which consituted the NESS using the same criterion described in Section 
\ref{s:results}. In the second procedure, the 
simulation was initialized, for a run at particular values $(P,H_b)$, 
such that the final saturation field before the measurement sequence was
of the same sign as $H_b$. This field-symmetric initial condition begins
each simulation near the NESS, for both positive and negative $H_b$, thus
eliminating the transitions seen in Fig.~\ref{mt.data.experiment}(a). 
We then ran 32 independent MC realizations of 50 field cycles each,
including the final 40 cycles of each realization in the NESS data. 
The simulated NEPDs of $\langle Q \rangle $ and $\sigma ^2 (Q)$ using the
first procedure are shown in Figs.~\ref{nepd.sim.singlerun}(a) and
\ref{nepd.sim.singlerun}(b), respectively, while those produced using the
second procedure are shown in Figs.~\ref{nepd.sim.steadystate}(a) and 
\ref{nepd.sim.steadystate}(b).

\begin{figure}[ht]
\begin{center}
\includegraphics[width=3.0in,angle=0]{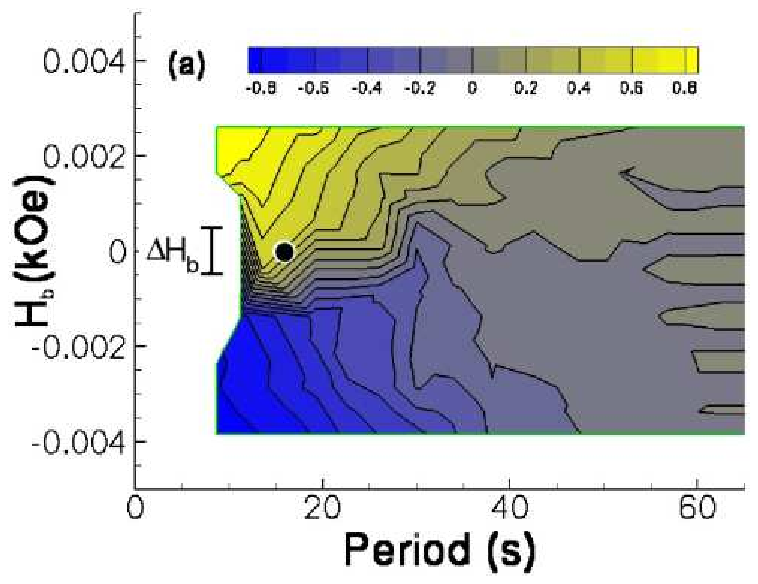} 
\includegraphics[width=3.0in,angle=0]{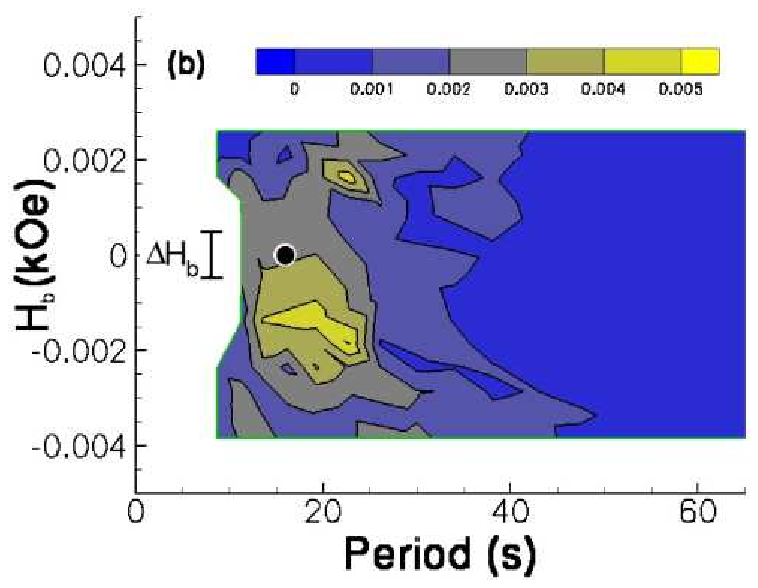} 
\end{center}
\caption
{\label{nepd.sampleA} (color online). 
Non-equilibrium phase diagrams (NEPDs) for sample A, showing (a) the average  
$\langle Q \rangle$ and (b) the variance $\sigma^2 (Q)$ in the non-equilibrium 
steady state (NESS), as functions of the period $P$ and bias field $H_b$.
The portion of the magnetization time series which consituted the NESS was determined 
for each ($P, H_b$) according to the procedure described in Section \ref{s:results}. Within 
each time series, the bias field fluctuated within a range 
$(H_b-\Delta H_b/2,H_b+\Delta H_b/2)$, with $\Delta H_b = 1.0$~Oe as shown. 
The black dot shows the estimated location of the critical point of the DPT.}
\end{figure}

We concentrate first on the comparison of simulation data to the data from
experimental sample A, for reasons discussed below. 
The simulated NEPDs of Fig.~\ref{nepd.sim.singlerun}
are clearly similar in form to the experimental NEPDs in 
Fig.~\ref{nepd.sampleA}. However, there are two significant differences, which
can be accounted for by the presence in sample A of 
hard-to-reverse, residual bubble domains, and of 
a pinning field for domain-wall motion, respectively. The first difference
is in the position of the peak of the variance $\sigma ^2 (Q)$. In the
experimental NEPD, the peak of $\sigma ^2 (Q)$ is situated at
($P \approx 20$~s$, H_b \approx -1.5$~Oe), whereas in the simulated NEPD, the
(somewhat diffuse) peak occurs at periods just above $P_c = 493$~MCSS. While one
might argue from Fig.~\ref{nepd.sim.singlerun} that the peak is
centered around a slightly negative $H_b$, a collection of NEPDs
(produced by different MC realizations) showed no net tendency toward either
positive
or negative $H_b$. As described at the end of Section \ref{s:results}, the 
shift in the boundary between the regions of positive and negative 
$\langle Q \rangle$ values, as well as the shift of the experimental peak 
of $\sigma ^2 (Q)$, toward negative $H_b$, 
can be explained by the asymmetry in nucleation 
fields during the measurement sequence. This asymmetry is in turn due to the
presence of the residual bubble domains, which were magnetized 
positively by the final (positive) saturating field before the beginning of the 
measurement sequence.

The second principal difference between the simulated and experimental
NEPDs is the gradual slope of the contour lines
extending from the left side of the experimental NEPD in
Fig.~\ref{nepd.sampleA}(a), as compared to the steep drop of the corresponding
contours in the simulated NEPD in Fig.~\ref{nepd.sim.singlerun}(a). If one 
normalizes the $H_b$ values (i.e., the $y$ axis) by the field magnitude $H_0$
in both Figs.~\ref{nepd.sampleA}(a) and \ref{nepd.sim.singlerun}(a),
this difference in slopes becomes even more pronounced. Physically, the
difference in slopes corresponds to a susceptibility 
$\partial \langle Q \rangle / \partial H_b$ which falls off more slowly with
increasing period in the experiment than in the simulation.
This increased sensitivity to the
bias field results from the pinning field in the multilayer. In the
experiment, the entire magnetization reversal takes place within the restricted
field intervals $(-H_0+H_b,-H_p)$ and $(H_p,H_0+H_b)$, 
whereas in the simulated reversal
it occurs, albeit with a domain wall velocity decreasing linearly with the
applied field, \cite{kn:rikvold00} 
within the entire field intervals $(-H_0+H_b, 0)$ and $(0, H_0+H_b)$.
A given bias field, expressed as a percentage of $H_0$, thus constitutes
a larger percentage of the field interval during which
magnetization reversal occurs in experiment than in the simulation, 
and so has a larger effect on $\langle Q \rangle$.

When the second data analysis procedure was used, including averaging over 32 independent
simulation runs, the NEPD of $\langle Q \rangle$, shown in Fig.~
\ref{nepd.sim.steadystate}(a), became symmetric with respect to $H_b$, both
above and below $P_c$. In addition, the NEPD of $\sigma ^2(Q)$ in 
Fig.~\ref{nepd.sim.steadystate}(b) became more sharply focused near 
$H_b = 0$ at periods just above $P_c$. 
This strongly suggests that the asymmetry with
respect to $H_b$ in Fig.~\ref{nepd.sampleA}(a), and size of the
peak region in Fig.~\ref{nepd.sampleA}(b), would decrease if
the second procedure described above were followed in gathering and
analyzing experimental data.

\begin{figure}[ht]
\begin{center}
\includegraphics[width=3.0in,angle=0]{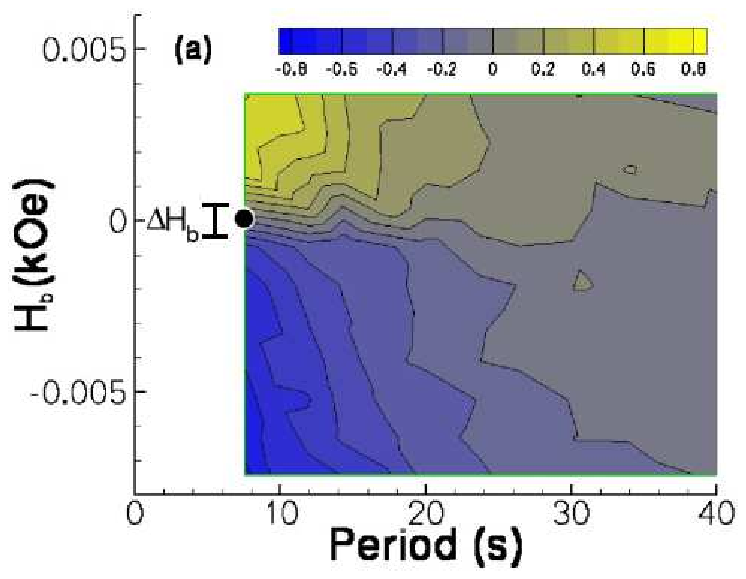} 
\includegraphics[width=3.0in,angle=0]{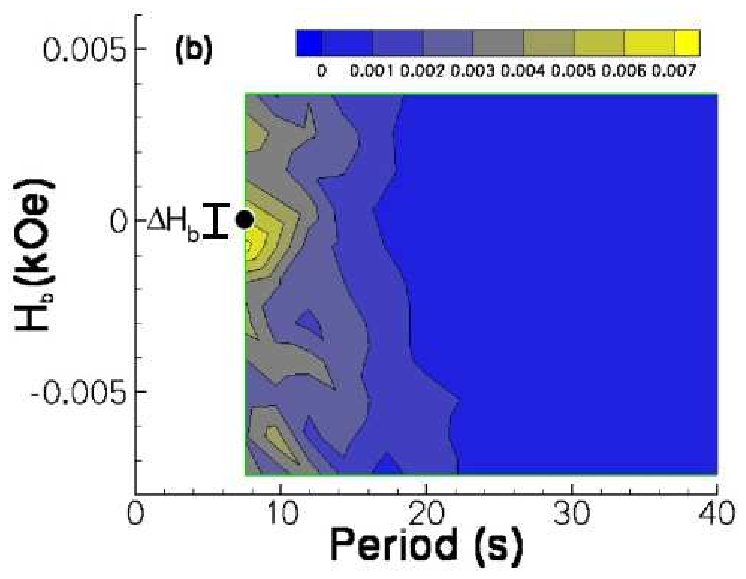} 
\end{center}
\caption
{\label{nepd.sampleB} (color online). 
Same as Fig.~\ref{nepd.sampleA}, but for sample B.}
\end{figure}

\begin{figure}[ht]
\begin{center}
\includegraphics[width=3.0in,angle=0]{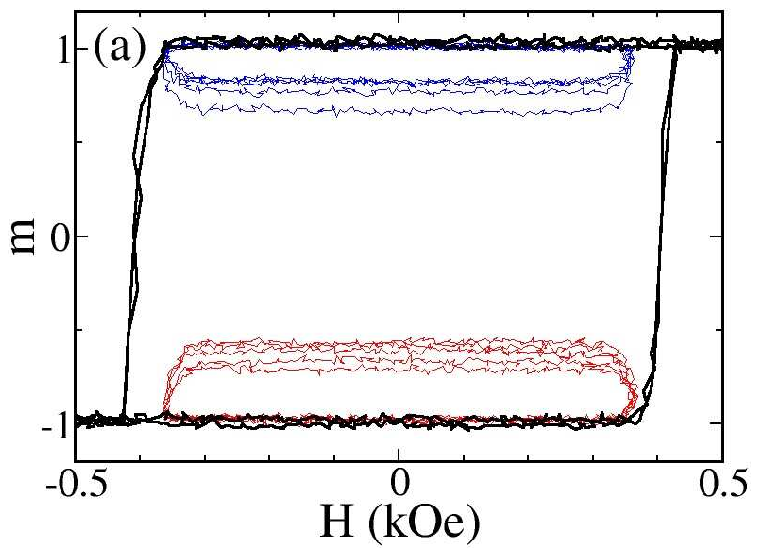} 
\includegraphics[width=3.0in,angle=0]{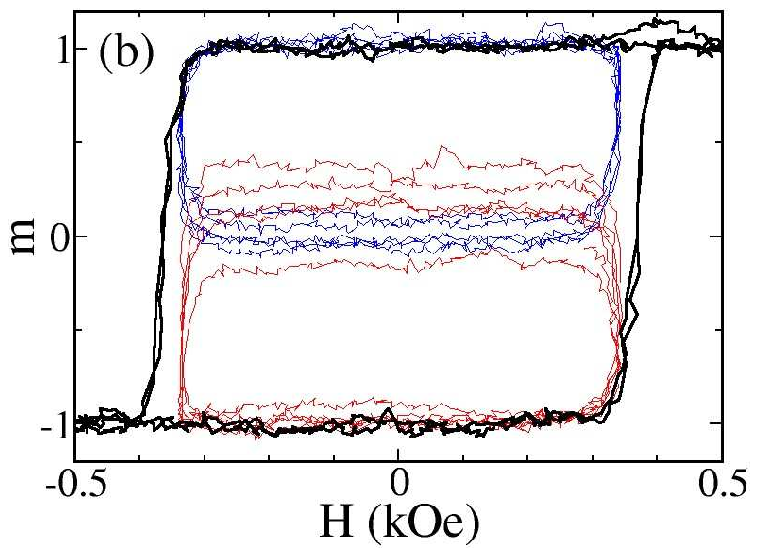} 
\end{center}
\caption
{\label{illustrate.persistent} (color online).
Normalized magnetization, $m$, vs field, $H$, in (a) 
sample A at ($P=8.7$~s, $H_b=-1.9\pm0.5$~Oe), and (b) sample B at
($P=7.6$~s, $H_b=-0.8\pm0.5$~Oe). The upper thin lines (blue) show the first 5
measurement cycles, which reach (nominal) positive saturation, while the
lower thin lines (red) show the final 5 measurement cycles, which reach
(nominal) negative saturation. The thick black lines show two complete
cycles at the saturation field, $H_s = 0.740$~kOe. The nucleation fields in
the measurement and saturated cycles correspond closely, except for
the increasing branch (from negative saturation) in (a), due to
the presence of positively magnetized residual bubble domains in sample A (see text). As the
$x$ axis range was reduced to more clearly show the nucleation fields, 
the saturation loops continue outside the field range shown.}
\end{figure}

\begin{figure}[ht]
\begin{center}
\includegraphics[width=3.0in,angle=0]{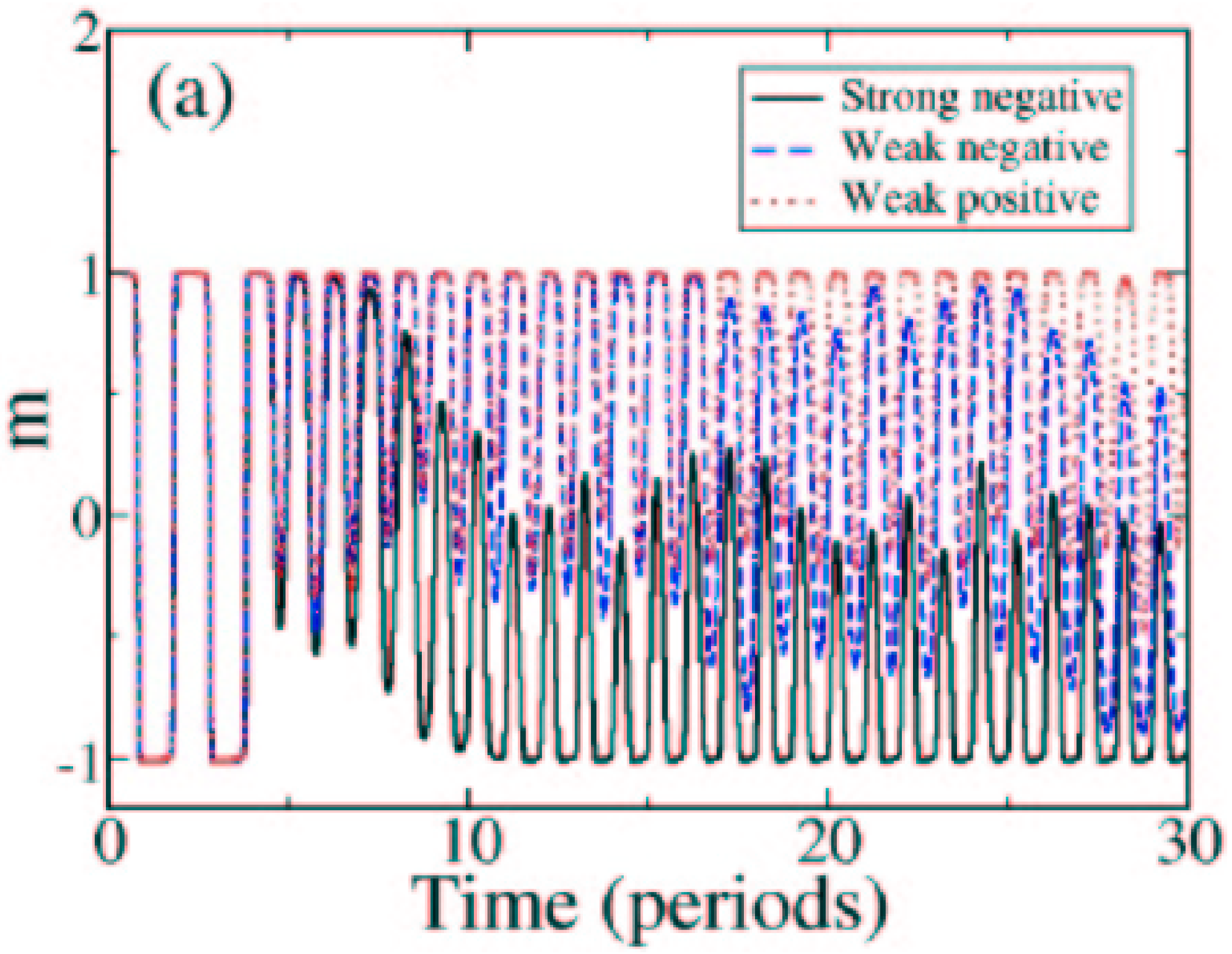} 
\includegraphics[width=3.0in,angle=0]{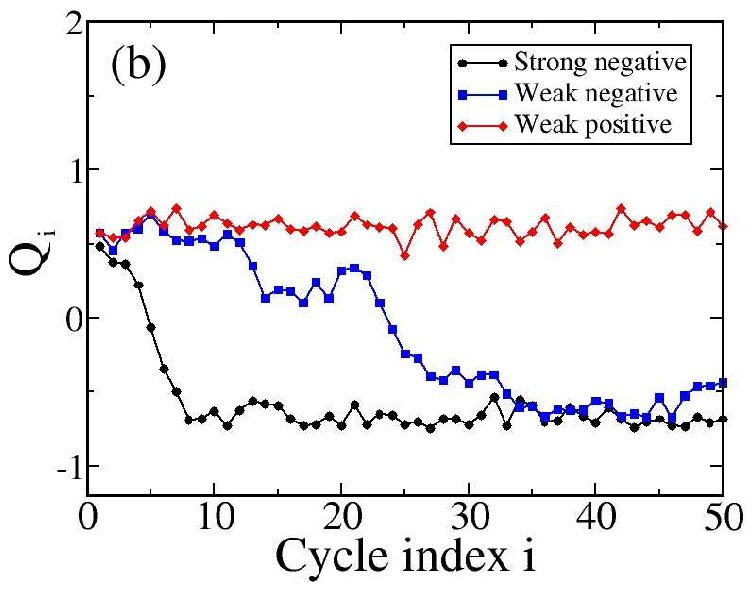} 
\end{center}
\begin{center}
\includegraphics[width=3.0in,angle=0]{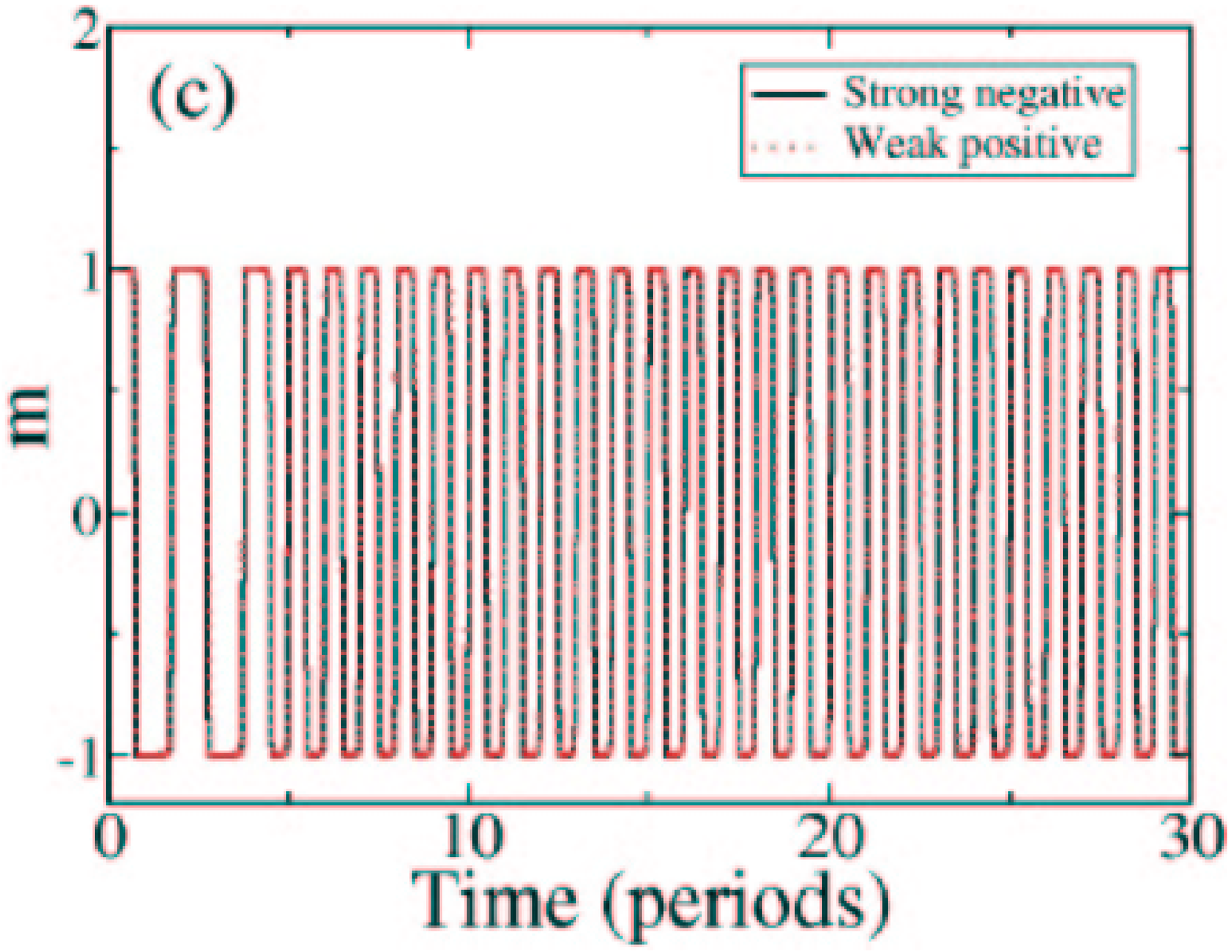} 
\includegraphics[width=3.0in,angle=0]{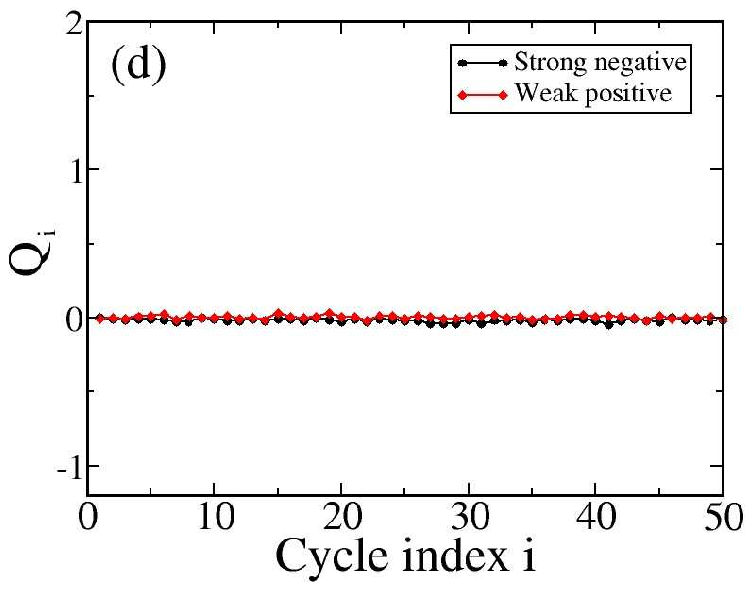} 
\end{center}
\caption {\label{mt.data.simulation} (color online).  
Time series of normalized magnetization, $m$, and of cycle-averaged magnetization, $Q_i$, from
simulations of the $L=180$ kinetic Ising model, at two different values of the period and 
in various bias fields. (a) Period $P = 473$~MCSS, magnetization time series. 
(b) $P = 473$~MCSS, $Q_i$ vs
measurement cycle $i$. (c) $P=1500$~MCSS, magnetization time series. (d) $P=1500$~MCSS, $Q_i$ 
vs measurement cycle $i$. As in Fig.~\ref{mt.data.experiment}, for clarity, only the two initial saturated cycles and the first
26 measurement cycles of the magnetization time series are plotted. However, the full time
series of $Q_i$, calculated from the 50 measurement cycles, are shown. 
The bias fields values corresponding to strong negative, weak negative, and weak positive are 
$-0.013J$, $-0.0035J$, and $+0.0044J$.} 
\end{figure}

\begin{figure}[ht]
\begin{center}
\includegraphics[width=3.0in,angle=0]{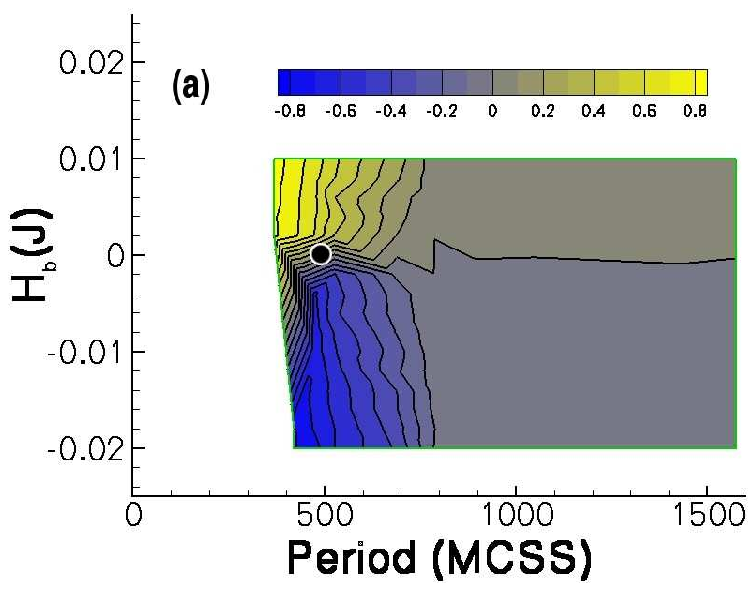} 
\includegraphics[width=3.0in,angle=0]{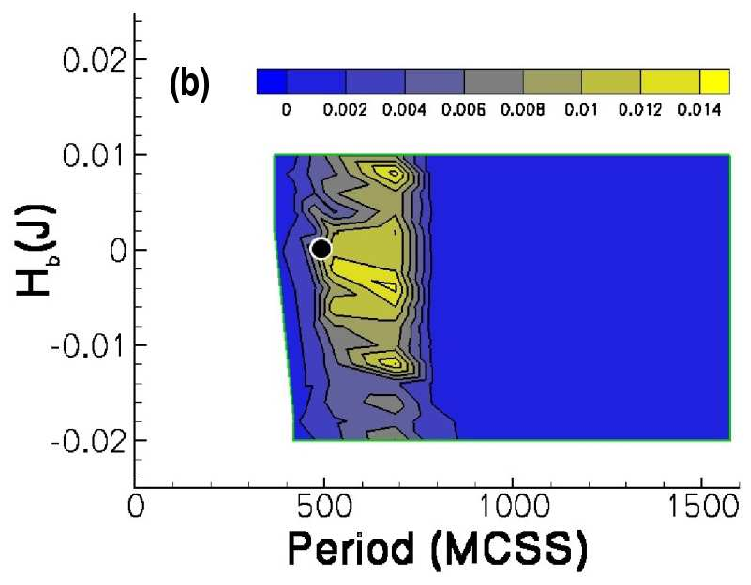} 
\end{center}
\caption {\label{nepd.sim.singlerun} (color online).  
Non-equilibrium phase diagrams (NEPDs) for the $L=180$ kinetic Ising model, 
with parameters given in Section \ref{s:comparison.discussion}, of 
(a) $\langle Q \rangle$ and (b) $\sigma^2 (Q)$ in a non-equilibrium 
steady state (NESS), as functions of the period $P$ and bias field $H_b$. 
The data for the NESS were drawn from a single run of 50 field cycles, using
the same analysis which was employed for the (experimental) 
Figs. \ref{nepd.sampleA} and \ref{nepd.sampleB} and described in Section
\ref{s:results}. The black dot shows the location of the critical point, 
$(P=493 \pm 2$~MCSS $, H_b=0)$, as determined by finite-size scaling analysis.}
\end{figure}

\begin{figure}[ht]
\begin{center}
\includegraphics[width=3.0in,angle=0]{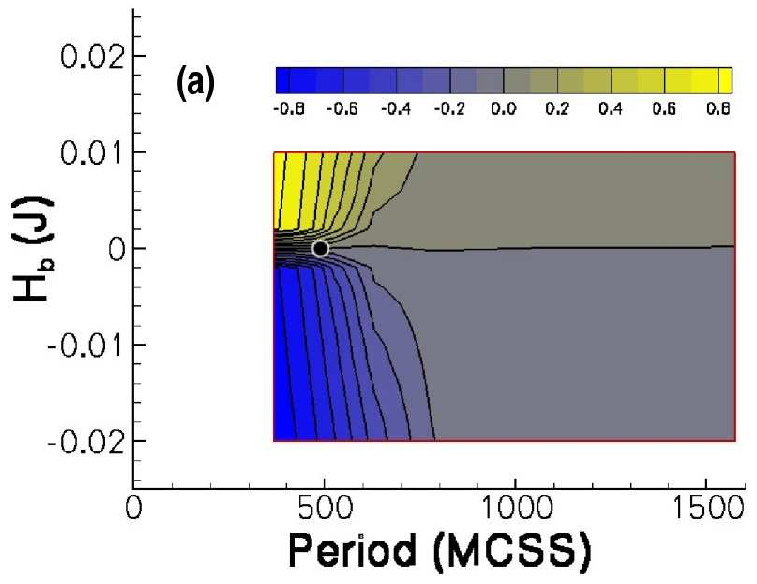} 
\includegraphics[width=3.0in,angle=0]{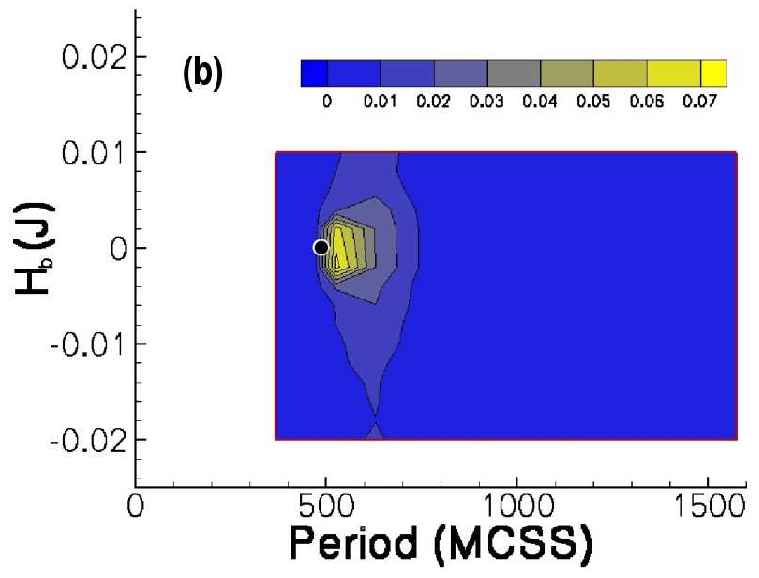} 
\end{center}
\caption {\label{nepd.sim.steadystate} (color online).  
Same as Fig.~\ref{nepd.sim.singlerun}, but using a different procedure to generate the data
in the non-equilibrium steady state (NESS). Here, the data for the NESS were drawn from 
the final 40 cycles and averaged over 32 independent
MC simulations, each of which was initialized in a saturation field of the
same sign as its associated $H_b$. }
\end{figure}

In comparing the simulation data to the results for experimental sample B,
we note that in the NEPD of $\langle Q \rangle$ for sample B in
Fig.~\ref{nepd.sampleB}(a), there is much less (if any) shift of the
boundary between the region of positive $\langle Q \rangle$ and the region
of negative $\langle Q \rangle$ toward negative bias field. Similarly, the peak
of $\sigma^2(Q)$ in Fig.\ref{nepd.sampleB}(b) is just slightly further
from $H_b = 0$~Oe than the uncertainty $\Delta H_b / 2 = 0.5$~Oe. The very
minimal difference between the nucleation field in the increasing saturating
fields and that in the increasing measurement loops in 
Fig.~\ref{illustrate.persistent}(b), as compared to the sizable difference
noted above for sample A in Fig.~\ref{illustrate.persistent}(a), suggests
strongly that there were significantly fewer impurities and/or variations in
film thickness giving rise to residual bubble domains in
sample B. The NEPDs of $\langle Q \rangle$ and $\sigma^2(Q)$ for sample B
closely resemble the part of the simulated NEPDs 
(Fig.~\ref{nepd.sim.singlerun}) above $P=500$~MCSS ($\approx P_c = 493$~MCSS).
Thus, we tentatively place the location of the critical period at
($P=7.6$~s,~$H_b=0.0$~Oe), i.e., at the left edge of the NEPDs, as indicated
in Fig.~\ref{nepd.sampleB}. This is also consistent with the fact that
the magnetization reversal reaches $m \approx 0$ for sample B at $P = 7.6$~s in 
Fig.~\ref{illustrate.persistent}(b), whereas it reaches only $|m|=0.7$ for sample A at 
$P = 8.7$~s in Fig.~\ref{illustrate.persistent}(a). In simulations, 
it has been observed \cite{kn:sides99} that the DPT occurs at a period where 
the magnetization reversal (from saturation) proceeds at least as far as $m = 0$. 

Since the data from sample A includes periods from
both below and above our critical period, $P_c \approx 20$~s, with
the variance $\sigma ^2 (Q)$ visibly decreasing in Fig.~\ref{nepd.sampleA}(b) 
both below and above this period, we have focused
on sample A as our primary evidence for the observation of the DPT.
The similarity of the experimental results from sample A to the simulated results,
with respect to (i) the behavior of the time series $Q_i$ in various $H_b$, in 
Figs.~\ref{mt.data.experiment} and \ref{mt.data.simulation}, (ii) the form 
of the NEPDs of $\langle Q \rangle$, in  Figs.~\ref{nepd.sampleA}(a) and 
\ref{nepd.sim.singlerun}(a), and (iii) most importantly,    
the existence of a well-defined peak in the NEPD of $\sigma ^2 (Q)$ 
in Fig.~\ref{nepd.sampleA}(b);  constitute strong evidence 
for this DPT in an experimental magnetic system. 
The differences between the experimental
and simulated NEPDs of $\langle Q \rangle $ and $\sigma ^2 (Q)$ can be 
consistently explained, for both samples A and B, 
in terms of the effect of hard-to-reverse
bubble domains in sample A, as described in Section \ref{s:results}, and the presence
of a pinning field for domain-wall motion, as described earlier in this section.

As mentioned in Section \ref{s:intro}, the kinetic Ising model we have used
does not give rise to a pinning field, and thus presents a simplified
model of the motion of domain walls in the [Co/Pt]$_3$-multilayer. Significant
progress has been made in using a Preisach-Arrhenius (PA) model of 
magnetic viscosity to describe the
thermally activated motion of domain walls over a distribution of free-energy
barriers associated with variations in film thickness, grain boundaries, and
impurities.\cite{kn:dellatorre02,kn:chen02,kn:dellatorre98} The PA model,
while quite useful to describe phenomenologically the decay of magnetization
on long time scales, does not directly 
incorporate the spatial dependence and cooperative nature 
of magnetization processes. 
In addition, recent work \cite{kn:ferre04,kn:lemerle98}
has shown that the dependence of the energy barriers (to wall motion) on 
the applied field deviates from the 
linear dependence \cite{kn:dellatorre02,kn:chen02,kn:dellatorre98} usually assumed 
in PA models. It appears likely that
incorporating both spatially varying magnetization, such as that occurring
in the MD reversal process (cf. Fig~\ref{kerr.image}), 
and a realistic description of a wide distribution
of energy barriers will require a computationally intensive, multi-scale
micromagnetic model, for which the DPT would first have to be established
computationally. We have therefore chosen to model the spatially
varying MD reversal process faithfully with a two-dimensional kinetic
Ising model, and to account for magnetic viscosity effects (i.e., the pinning
field) in our analysis and comparison to experiment. The differences between
the results of the experiments and the kinetic Ising simulations are 
consistent with what one would expect from the added effects of magnetic viscosity 
in an experimental multilayer system exhibiting the DPT.

\section{Conclusion \label{s:conclusion}}

We have compared the behavior of the cycle-averaged
magnetization, $Q$, in experiments on a [Co/Pt]$_3$ multilayer, to its
behavior in MC simulations of the two-dimensional kinetic Ising model, in
which a DPT with respect to the period $P$ of an applied alternating field has
been well established. \cite{kn:korniss00,kn:sides99} Plots of time series of 
the magnetization (and of $Q$) in the presence of various bias 
fields $H_b$, as well as non-equilibrium phase diagrams of the average,
$\langle Q \rangle$, and the variance, $\sigma ^2 (Q)$, 
as functions of $P$ and $H_b$, were used to characterize and compare 
the behavior of $Q$ in experiment and simulation. 
The behavior was seen to be similar, with 
differences that could be clearly accounted for in terms of the known phenomena
of a pinning field and residual bubble domains.
The results, in particular the presence of a clear
peak in the variance $\sigma ^2 (Q)$ in experimental sample A,  
provide the strongest experimental evidence to date 
for the presence of this DPT in a magnetic system. 
Furthermore, the results strongly suggest that the cycle-averaged magnetization $Q$ serves as a dynamic order parameter in the [Co/Pt]$_3$-multilayer system.

It would be most desirable to perform further and more precise experiments on the DPT in perpendicular magnetic ultrathin films. Specifically, it would be useful to collect data using the second procedure described in Section 
\ref{s:comparison.discussion}, i.e., to ensure that the final saturated state before the measurement cycle has the same sign as the applied bias field. Also, one could attempt to apply a field magnitude $H_0$ that is significantly larger than the pinning field $H_p$, thereby forcing the critical period to lower values. One would expect the effect of the pinning field, for example in creating more gradual contours in the non-equilibrium phase diagram of 
Figs.~\ref{nepd.sampleA}(a) and \ref{nepd.sim.singlerun}(a), to be reduced in this case. However, this will be experimentally very challenging given the hysteretic nature and response time of the electromagnet one would need in this case to generate the time-varying applied field sequences.

In addition, a lower value of the critical period would enable the collection of data for more field cycles in each NESS. As noted above, some level of experimental fluctuations in the bias field is probably unavoidable. However, with a sufficiently large number of cycles recorded in a NESS, one could attempt to calculate values of $\langle Q \rangle$
and $\sigma ^2 (Q)$ using data from a subset of the recorded cycles with a more restricted range of $H_b$ values. This could potentially supply the precision required to investigate power-law scaling near the critical period. The added statistics provided by forcing the DPT to occur at a shorter period, as well as more precise control over the temporal
stability of the bias field, could enable extraction of the critical
exponents from the power-law behavior of $\sigma ^2 (Q)$ vs
$P - P_c$ and $H_b$.

Finally, in an ultrathin Au/Co/Au film it was 
observed \cite{kn:kirilyuk97} that a mixture of nucleation at fixed (`soft', extrinsic) sites and at random (intrinsic) sites occurred. 
If this also occurs in [Co/Pt] multilayers, 
it could affect both the size of the fluctuations of the 
dynamic order parameter and its spatial correlations, relative to those in the 
kinetic Ising system, in which nucleation occurs only at random sites. 
One could examine a series of real-time images of the sample, such as the one shown in Fig.~\ref{kerr.image}, in order to calculate spatial and time-displaced 
correlation functions to address this question. 

The results from such additional experiments could serve to
further reinforce the 
evidence for the DPT presented here. To achieve an even higher degree of
confidence, one would
need to formulate an accurate, multiscale micromagnetic model of the 
multilayer, and to demonstrate the existence of the DPT in this computational
model. Given the large system sizes needed for computational 
finite-size scaling analysis, such an effort will likely require 
innovations in simulation methods, as well as further increases in available 
computing power.

\section{Acknowledgements}
The computational work was supported by NSF Grant 
Nos. DMR-0120310 and DMR-0444051, by the Center for
Computational Sciences at Mississippi State University, and by the School of Computational Science at Florida State University. In addition, one of us (D.R.) acknowledges support from Clarkson University and the use of resources under NSF Grant No. DMR-0509104. Another author (A. B.) would like to acknowledge funding from the Department of Industry, Trade and Tourism of the Basque Government and the Provincial Council of Gipuzkoa under the ETORTEK Program, Project IE06-172, as well as from the Spanish Ministry of Science and Education under the Consolider-Ingenio 2010 Program, Project CSD2006-53.

%\bibliographystyle{plain}
%\bibliography{refs.dpt}

\begin{thebibliography}{34}
\bibitem{Ising.simplification.endnote} The kinetic Ising model does simplify certain aspects of the magnetization dynamics in the
multilayer, mainly the dependence of the domain-wall velocity on the applied field. This issue is discussed in detail in Section 
\ref{s:comparison.discussion}.
\bibitem{zero.hall.endnote}
The Hall probe was calibrated accurately by determining the reading of the 
Hall probe corresponding to true zero field at the sample as the point 
at which the cycle-averaged magnetization changed from negative to 
positive value at high field period.
\bibitem{pinning.method.endnote}
The pinning field was determined by first averaging 
over a collection of increasing (decreasing) $M-H$ curves, 
such as the increasing curves near 
$m=0$ in Fig.~\ref{mh.data}(a). The widest field range $(-H_{p,-},H_{p,+})$ for
which all $m$ values outside the field range were greater (smaller) than the
average value, $\langle m \rangle$, within the field range was then found. Finally, 
the endpoints $H_{p,-}$ and $H_{p,+}$ from eight such procedures were 
used to calculate the mean value and uncertainty of $H_p$ given in the text.
\bibitem{no.scaling.endnote} 
Despite applied field magnitudes and bias field magnitudes
precise to the 0.5\% and 0.1\% level, respectively, of the applied field
magnitude, it was not possible to establish scaling relations with statistical
significance from our data.
\end{thebibliography}

\newpage

\end{document}